# Ambiguous Red Shifts

Carl E. Wulfman
Prof. Emeritus, Dept. of Physics,
University of the Pacific, Stockton, CA, U.S.A
email: wulfmanc@wavecable.com

*Dedicated to the memory of Marcos Moshinsky who contributed so much to science and provided so many opportunities to so many*

A one-parameter conformal invariance of Maxwell's equations allows the wavelengths of electromagnetic waves to change as they propagate, and do so even in otherwise field-free space. This produces an ambiguity in interpretations of stellar red shifts. Experiments that will determine the value of the group parameter, and thereby remove the ambiguity, are proposed. They are based on an analysis of the anomalous frequency shifts uncovered in the Pioneer 10 and 11 spacecraft studies, and physical interpretation of an isomorphism discovered by E. L. Hill. If the group parameter is found to be non-zero, Hubble's relations will have to be reinterpreted and space-time metrics will have to be altered. The cosmological consequences of the transformations are even more extensive because, though they change frequencies they do not alter the energy and momentum conservation laws of classical and quantum-electrodynamical fields established by Cunningham and by Bialynicki-Birula.

## 1. Introduction

Four years after Einstein introduced special relativity, Bateman[1] and Cunningham[2] proved that Maxwell's equations are invariant under space-time inversions, and under the transformations of a fifteen-parameter conformal Lie group which contains the Poincare group. Just as WW II ended, E. L. Hill[3] pointed out that one of its transformations, with Lie generator

$$C = (r^2/2c^2 + t^2/2)\partial/\partial t + t\, r\partial/\partial r = C_4/2c,$$

$$C_4 = (r^2 + x^{4\,2})\partial/\partial x^4 + 2x^4\, r\partial/\partial r, \quad x^4 = ct, \tag{1}$$

establishes a relation isomorphic to Hubble's law. Though Kastrup[4] lists Hill's paper in a review of conformal invariance, it appears that for 65 years Hill's

(1)

discovery itself led to no published analyses. Recently though, it has been shown that Hill's transformation alters both metrics and wave-lengths of propagating EM waves.[5] As will be seen below, this has the consequence that the physical interpretation of Hill's isomorphism requires considerable care.

Though the isomorphism is non-relativistic, a number of author's have used the Bateman-Cunningham group in relativistic studies of Hubble's relations. The full Lie group was used by Hoyle, Burbidge, and Narlikar[6] to develop a generalized relativistic cosmology, and Bunn and Hogg[7] have used relativistic parallel-transport arguments to give kinematic treatments of Hubble's relations. In a quite general investigation of the Bateman-Cunningham Lie group, L. M. Tomilchik[8] argued that the anomalous frequency shifts uncovered in the Pioneer 10 and 11 programs imply that Hubble's relations are in fact the result of coordinate transformations of the group. Turyshev and Toth[9] subsequently provided a thorough analysis of the Pioneer studies which is partly responsible for the somewhat different conclusion developed below.

   Cunningham[2] proved that the operation of inversion, which alters spacetime metrics, does not alter the energy of the electromagnetic field in the region of length $\lambda$ that is bounded by planes normal to the propagation vector of a wave of wavelength $\lambda$. The same is true of the operations of the special conformal group, which also alter spacetime metrics. As Einstein's relation $E = h\nu$ requires that as $\lambda$ increases, the energy and momenta of photons decreases, it might appear that Cunningham's observation does not carry over into quantum mechanics. However, an analysis by Bialynicki-Birula[10] shows that when the appropriate scalar product is used in quantum electrodynamics, the special conformal transformations do not alter these conservation laws.

   It is apparent that Hill's isomorphism has so many implications that it would be hard to exaggerate the importance of carrying out experiments which could establish whether EM waves do behave as Hill's equation requires. As noted in Section 3, below, it is now possible to carry out experiments more definitive than those of the Pioneer program, experiments that could determine a reliable value of the group parameter. This would securely establish the rate of change, zero or otherwise, of the frequency of propagating EM waves.

                                        (2)

## 2. The Conformal Transformation

   The inversion transformation, $I_k$ , of the Bateman-Cunningham group may be defined by the relation

$$\mathbf{X} = I_k\mathbf{x} = k^2\mathbf{x}/\mathbf{x}\cdot\mathbf{x}, \quad \mathbf{x} = (x^1, x^2, x^3, x^4), \qquad (2)$$

in which $k^2 > 0$, and

$$\mathbf{x}\cdot\mathbf{x} = g_{ij} x^i x^j = x^{42} - r^2, \quad r = |(x^1, x^2, x^3)|, \; x^4 = ct. \quad (3)$$

The inversion is self inverse: $I_k = I_k^{-1}$, and it leaves invariant the angles

$$\phi = \arccos(\mathbf{x}\cdot\mathbf{x}'/|\mathbf{x}||\mathbf{x}'|). \qquad (4)$$

The similarity transformations produced by $I_k$ do not alter the generators of the Lorentz group, but convert the generators, $\partial/\partial x^a$ of translations of the variables $x^a$ to generators, $C_a$, of the special conformal group acting on the variables $x^a$. Because the $C_a$ do not commute with the spacetime translation generators, translation invariance in the one system of coordinates does not correspond to translation invariance in the other.

If $J_n$, $n = 1..10$, denotes generators of the usual realization of the Poincare group, the operators $J'_n = \exp(\beta C_4) J_n \exp(-\beta C_4)$ are generators of another realization of the group. The invariant angles between vectors **x** and **x**' defined by eqn (4) are not altered by this change of realization.

In the following discussion, k is set equal to 1. The operator $\exp(\beta C_4)$ then produces mappings between ordinary, and conformal, time variables, t, t', and radial position variables, r, r'. From Barut, Raczka[11] one has

$$r \to r' = \gamma r, \quad x^4 \to x'^4 = \gamma (x^4 - \beta s^2). \qquad (5)$$

Here

$$\gamma = (1 - 2\beta x^4 + \beta^2 s^2)^{-1}, \quad s^2 = (x^4)^2 - r^2. \qquad (6)$$

It follows that

$$dr \to dr' = \gamma^2 (A\, dr + B\, dx^4), \quad dx^4 \to dx'^4 = \gamma^2 (B\, dr + A dx^4), \qquad (7)$$

with

$$A = 1 - 2\beta x^4 + \beta^2 (r^2 + (x^4)^2), \quad B = 2\beta r (1 - \beta x^4). \qquad (8)$$

The transformations generated by $1 + \alpha C$ suffice for our purposes. They are:

$$(3)$$
$$r \to r' = (1 + \alpha t)r, \quad t \to t' = t + \alpha (t^2 + r^2/c^2)/2. \qquad (9)$$

and

$$J_k \to J'_k = J_k + \alpha [C, J_k]. \qquad (10)$$

If v' = dr'/dt', and v = dr/dt, then to first order in $\alpha$ and v/c, eqns. (7), (8) yield Hill's relation

$$v' = v + \alpha r, \qquad (11)$$

isomorphic to Hubble's relation, eq. (21), below.

Two further properties of the transformations produced by C and $C_4$ are required to physically interpret the relations they produce.

i) They transform metrics. For example, if T represents time measured from the origin of a light cone, Minkowski's metric $ds^2$ is transformed to

$$ds'^2 = \gamma^2 ds^2 \to (1 + 2\alpha T)ds^2 + O(\alpha^2), \qquad (12)$$

Thus Hill's equation establishes a relation between velocities measured in spacetimes with two different metrics. The full transformation, defined by eqns. (7) and (8), does not introduce any frequency dependence into metrics.

ii) Using eqns. (7), (8) to determine the times $\Delta T$, $\Delta T'$ that it takes a single cycle of a wave to pass a point R, R', one finds that EM waves with local wavelength $\lambda$ are transformed into waves with local wavelengths

$$\lambda' = \lambda\{(1+ \alpha T + (3/4)\, \alpha^2(t^2 + r^2/c^2) + O(\alpha^3) \}. \qquad (13)$$

Thus to $O(\alpha)$,

$$\lambda' = \lambda(1 + \alpha T) = \lambda(1 + \alpha T'). \qquad (14)$$

The wave-lengths $\lambda$, $\lambda'$ are those of the single wave-cycle centered at a point P with coordinates R ,T, R',T'. For our purposes one may consider the source of the wave to be at the origin, at which $\lambda' = \lambda$. Doppler shifts in $\lambda'$ correspond to source – observer velocities $V'(T') = c\Delta\lambda'/\lambda'$ in a spacetime with metric $ds'^2$. If $V/c <<1$, then in agreement with eq. (11), the change in $\lambda'$ at P requires that to $O(\alpha)$

$$c\Delta\lambda'/\lambda' - V = V' - V = \alpha R = \alpha R'. \qquad (15)$$

(4)

## 3. A First Estimate of The Value Of The Group Parameter

In this and the following Section the results of the Pioneer 10 and 11 programs are analyzed and used as the basis for proposing measurements that should be able to determine the value of the group parameter $\alpha$ with an accuracy limited only by the accuracy which the necessary Doppler measure ments can attain. The analysis is made possible by the work of Turyshev and Toth[9], who have thoroughly investigated the records of the Pioneer 10 and 11 spacecraft experiments first described by Anderson, et al[12]. In most of these experiments, pulses of ~13.6cm S band radar waves were sent from earth to the receding spacecraft, which returned amplified signals that had been phase-locked at the spacecraft to the signal it received. A microwave interferometer was used to measure Doppler shifts with an accuracy limited only by thermal or plasma noise. This made it possible to determine mean radial velocities $V_{Dop}$ of Pioneer 10 in a 60 sec., or greater, intervals to within 0.3 mm/sec.

As the return times of the signals were not measured, a sophisticated dynamical model was used to predict spacecraft position and velocity vectors and velocities relative to earth-based interferometers. The dynamical model assumed the spacetime metric to be of a parametrized post-Newtonian form accurate to $O((v/c)^4)$, and the equations of motion took into account a variety of further effects. Velocities determined from observed Doppler shifts in radar frequencies took into account relativistic effects on the propagation of the radar waves, and on time standards, and the effects of propagation through the earth's atmosphere. Never the less, as Pioneers 10 and 11 receded beyond 20 AU, the predicted and observed velocities developed regularly increasing small differences. These systematic differences, the "Pioneer anomaly", were originally expressed as a function of the distance from the spacecraft to the sun, but periodic oscillations indicate that they are better expressed in terms of radial distances

R to the earth, and radial speeds $V_{pred}$ and $V_{Dop}$. The data establish that between 20 AU and 70 AU,

$$V_{pred} - V_{Dop} = -a\,R, \quad a = (2.84 \pm \varepsilon) \times 10^{-18} \text{sec}^{-1}, \quad \varepsilon < 0.42. \quad (16)$$

Eq. (16) expresses results from both spacecraft, but only signals from Pioneer 10 were usable through the entire interval. It reached 70 AU eight years after it was launched. The observed anomalous wavelength changes varied from approximately $(4 \pm 0.3)$mm to $(14 \pm 0.3)$mm. For the values of R in eq. (16)

$$(5)$$

it was supposed that the metric was well approximated by that of Minkowski. However, if EM waves travel in spacetime with the metric $ds'^2$, then $V_{Dop}$ in eq. (16) approximately equals $V'$ in eq. (15). As $\lambda_{pred}$ was obtained on the assumption it is an ordinary wavelength $\lambda$, the $V_{pred}$ in eq. (16) corresponds to V in eq. (15). It follows that

$$V' - V = a\,R. \quad (17)$$

The anomaly then becomes a physical red shift rather than the blue shift of eq. (16). As the waves traveled a distance 2R, eqn's (15) and (17) would imply that $\alpha = a/2$. However as Turyshev and Toth emphasize, the dynamical model which produced the predicted motions is now the subject of investigations which may considerably alter the value of $a$. Despite this uncertainty, the Pioneer observations establish the significant fact that it is possible to measure values of $\alpha$ of the order of $10^{-18}$sec$^{-1}$ with an accuracy about the same as that to which (see below) Hubble's constant is known.

## 4. On the Experimental Determination of the Value of the Group Parameter

The dependence of the Pioneer results upon modeling of the dynamics of spacecraft motion can be removed by directly measuring the times taken by transmitted radar signals to return from the spacecraft, and using these times to calculate its positions and velocities. The parameter $\alpha$ can be most simply calculated if it is arranged that during a series of measurements of radar-pulse return times, $T_{ra}$, the radial distance $R_{ra} = cT_{ra}$ to a spacecraft or a distant satellite, does not change. If the spacetime metric is $ds'^2$, the Doppler measurements of the radial velocity $V'_{Dop}$, will satisfy the relations

$$\Delta\lambda_{Dop}/\lambda_{Dop} \;\square\; \alpha T_{ra}. \quad (18)$$

When the radar pulse delay time measurements show that the distance to the spacecraft is changing, they can be used to calculate its velocities $V_{ra}$. A series of such measurements will then produce the generalization of eq. (18):

$$\Delta\lambda_{Dop}/\lambda_{Dop} = V_{ra}/c + \alpha T_{ra}. \quad (19)$$

Subtracting the first term on the RHS enables one to calculate $\alpha$ as before. In

<div align="center">(6)</div>

both cases, if the waves do not elongate as they propagate, it will be found that the group parameter vanishes.

Cesium clocks now provide time standards accurate to 3 parts in $10^{16}$ (Petty[13]). This establishes the limit on the accuracy of measurements of $T_{ra}$ in eqns. (18) and (19), and a limiting accuracy of 6 parts in $10^{16}$ on measurements of the first term on the RHS of eq. (19). At a typical spacecraft velocity of 14 km/sec $= 4.67 \times 10^{-5}$ c, the uncertainties in Doppler measurements of velocities correspond to relative accuracies of the order of 1 part in $4.5 \times 10^{7}$. Thus one can expect that it is the Doppler measurements that will limit the accuracy with which $\alpha$ can be determined from eqns. (18) and/or (19).

To determine whether neglected terms of $O(\alpha^2)$, $O(v^2/c^2)$ could alter this conclusion, we note that eqns. (7), (8) imply that on setting r = R, t = T, and v = V , and including the higher order terms in eq. (11) one has

$$\alpha R \ \to\ \alpha R(1 - V^2/c^2) + \alpha R \, ( \, \alpha R /c)(1 - V^2/c^2)(1/2 - V/c) + O((\alpha R)^3). \quad (20)$$

Suppose that R is large, say $1.5 \times 10^{8}$ km, (about 70AU), that v/c $= 10^{-4}$ and that $\alpha = \kappa \times 10^{-18}$ sec$^{-1}$, with $-10 < \kappa < 10$. Then $\alpha R$ is $1.5 \, \kappa \times 10^{-10}$ km/sec , $\alpha R/c$ is $5 \, \kappa \times 10^{-16}$, and $(1 - V^2/c^2)$ is $(1 - 10^{-8})$. On inserting these values into eq. (20) it becomes evident that the value of $\alpha R$ is most affected by the relativistic correction in the first term, (as is the Doppler shift). One concludes that the change in $\alpha R$ , and hence $\alpha R'$ and $\alpha R_{ra}$, produced by neglecting the higher order terms is less than the accuracy to which the Doppler shift can be measured. One obtains the same result by inserting appropriate values of r, $x_4$, and $dr/dx_4$ directly into the ratio $dr'/dx_4'$ defined by eqns. (7), (8). Comparison of the stability of the measurements over periods of 10 years leads to the same conclusion. It is apparent that the limitations on the accuracy of the determination of $\alpha$ can be expected to be the limitations on the accuracy of Doppler measurements that were found in the Pioneer program. Thus it should be possible to determine $\alpha$ with an accuracy of at least $\pm \Box\Box \times 10^{-18}$sec$^{-1}$.

### 5. Physical Interpretation of Hill's Isomorphism

In Hubble's equations,
$$V_{Hub} = V + H_0 R, \qquad\qquad\qquad (21)$$

<div align="center">(7)</div>

$V_{Hub}$ is the radial velocity of a star with respect to an observer, $H_0 R$ is a nearly universal rate of recession of nebulae ("expansion rate of the universe"), and V is the radial component of the local velocity of the star with respect to the mean radial velocity of the stars in a nebula. $V_{Hub}$ is determined from observed

Doppler shifts and the equation

$$c \, \Delta\lambda/\lambda = V_{Hub}. \qquad\qquad (22)$$

If no very strong gravitational fields need be taken into account, it may be assumed that R and V, and $V_{Hub}$ are measured in spacetime with Minkowski metric. Peebles[14] summarizes the evidence for Hubble's relations, and finds the value of $H_0$ to be $(2.19 \pm 0.56) \times 10^{-18} sec^{-1}$. Interpretations of both the redshift and the Cepheid luminosity evidence underpinning Hubble's equations depend upon the presupposition that EM waves behave as usually assumed.

If it is found that the spacetime metric is a conformal one, then the measurements leading to eq. (21) must be considered measurements of conformal coordinates and velocities. This will establish the physical meaning of the isomorphism between Hill's and Hubble's equations, and establish that only a portion of $H_0$ represents a general expansion rate. Thus, one may write $H^0 = h_0 + \alpha$, and rewrite eq. (21) as

$$V'_{Hub} = V' + (h_0 + \alpha)R'. \qquad\qquad (23)$$

## 6. Concluding Remarks

The previous discussion has investigated a current uncertainty in the behavior of propagating electromagnetic waves, their field energy and momentum, and the metric of the spacetime in which they propagate. The uncertainty is produced by a lack of knowledge of the parameter $\alpha$ in the operator $\exp(\alpha C)$ which can elongate propagating waves and change metrics, while leaving Maxwell's equations and field energies and momenta invariant in a well-defined sense. Though the Pioneer 10 and 11 studies provide an uncertain value of the group parameter $\alpha$, they provide a very useful body of experience. It has been used here as the basis for suggesting experiments that could reliably determine the value of $\alpha$. If $\alpha$ is found to have a nonzero value it will be necessary to alter interpretations of the values of the temporal and spatial coordinates of distant events. If it turns out that $\alpha$ has a value of the order of magnitude of Hubble's constant, interpretations of many astronomical

(8)

observations will have to be very greatly revised. All stellar Doppler shifts will have to be reinterpreted, and relations between them and stellar magnitudes will require alteration. Many cosmologies will be profoundly affected. Those dealing with failures of the conservation laws applicable to electro-magnetic fields will have to be revised to take into account the invariance of these laws under Bateman–Cunningham transformations.

If experiments establish that, contrary to the provisional results of the Pioneer studies, $\alpha$ is zero, none of these modifications will be required.

**Acknowledgements**

It is a pleasure to thank Jerry Blakefield, Viktor Toth, and Professors Tomilchick and Bialynicki-Birula for helpful discussions, and Bernardo Wolf
For help in preparing the manuscript.

## References

[1] H. Bateman, 1909 The Transformation of the Electrodynamical Equations
    *Proc. London Math Soc.* **7**, Series 2, 70 ; *ibid*, 223

[2] E. Cunningham, The Principle of Relativity in Electrodynamics and   an
    Extension Thereof,  *Proc. London Math Soc.* **8**, Series 2, 77 (1909)

[3] Hill, E. L., On a Type of Kinematical "Red-Shift", *Phys. Rev.* **68**, 232 (1945)

[4] H. A. Kastrup, On the Advancements of Conformal Transformation and
    their Associated Symmetries in Geometry and Theoretical Physics,
    Ann. d. Physik **17**, 631 (2008)

[5] C. E. Wulfman, *Dynamical Symmetry*, (World Scientific Publ. Co., Singapore,
    2010) Chapter 14

[6] F. Hoyle, G. Burbidge, N. Narlikar, *A Different Approach to Cosmology*,
    Cambridge, 2000

[7] E. Bunn, D. Hogg, The kinematic origin of the cosmological redshift,
    *Am. J. Phys.* **77**(2009)

(9)

[8] L. M. Tomilchik, The Hubble law as a kinematic outcome of the space-time
    conformal geometry. *AIP Conference Proceedings* Vol.**1205**,177(2010)

[9] S. G. Turyshev, V. Toth, The Pioneer Anomaly, *arXiv* 1001.3686v1
    (2010)

[10] I. Bialynicki-Birula, The photon as a quantum particle. *Acta Physica
    Polonica* **B37**, 935(2006)

[11] A. O., Barut, R. Raczka, *Theory of Group Representations and
    Applications*, 2$^{nd}$ Rev. Ed., Chapter 13, Sec. 4.

[12] J. D. Anderson, et al, Study of the anomalous acceleration of Pioneer 10
    and 11. *Phys. Rev.* **D65**, 082004

[13] E. M., Petty, *NIST Time and Frequency Bulletin 633* (2010),  Boulder, CO
    80305-3328, USA
.

[14]  P. J. E. Peebles, *Principles of Physical Cosmology*, Princeton, (1993)
      pp.  24-25, 70 –74, 106, 298

(10)